\documentclass[12pt]{article}
\usepackage[T2A]{fontenc}
\usepackage[utf8]{inputenc}
\usepackage{setspace,amsmath}
\usepackage{bbold}
\usepackage{amsfonts}
\usepackage{amssymb}
\usepackage{amsthm}
\usepackage{authblk}
\usepackage{cite}
\usepackage{fontenc}
\usepackage[dvips]{graphicx}
\usepackage[unicode,linktocpage=true,plainpages=false,pdfpagelabels=false]{hyperref}
\usepackage{bm}
\newtheorem{theorem}{Theorem}[section]
\newtheorem{corollary}{Corollary}[theorem]
\newtheorem{lemma}[theorem]{Lemma}

\begin{document}
\title{Heat kernel: proper time method, Fock-Schwinger gauge, path integral representation, and Wilson line}
\author[1]{A.~V.~Ivanov\thanks{E-mail: regul1@mail.ru}}
\author[2]{N.~V.~Kharuk\thanks{E-mail: natakharuk@mail.ru}}
\affil[1]{\it{St. Petersburg Department of Steklov Mathematical Institute of Russian Academy of Sciences, 27 Fontanka, St. Petersburg 191023, Russia}}
\affil[2]{\it{ITMO University, St. Petersburg 197101, Russia}}
\date{\vskip 5mm}
\maketitle
\begin{flushright}
Dedicated to the memory of L. D. Faddeev
on the occasion to his 85th anniversary
\end{flushright}
\vskip 10mm
\begin{abstract}
The proper time method plays an important role in modern mathematics and physics. It includes many approaches, each of which has its pros and cons. This work is devoted to the description of one model case, which reflects the subtleties of construction and can be extended to a more general cases (curved space, manifold with boundary), and contains two interrelated parts: asymptotic expansion and path intergal representation. The paper discusses in details the importance of gauge conditions and role of the ordered exponentials, gives the proof of a new non-recursive formula for the Seeley--DeWitt coefficients on the diagonal, as well as the equivalence of the two main approaches using the exponential formula.
\end{abstract}

\newpage
\section{Introduction}
The proper time method has been developing for more than eighty years and today it is an essential tool in theoretical and mathematical physics. This approach was first proposed by Fock \cite{1} in 1937 when working with the Dirac equation. Only 15 years later after the formulation of quantum field theory in covariant form, the proper time method appeared in the works of Nambu \cite{4} for the construction of the Green function for the Dirac  equation, and Schwinger \cite{3} in the study of gauge invariance. Since then, the approach has been actively used in theoretical physics. \\

\noindent The next important step in the development of the method is the DeWitt works \cite{i,i+1,i+2,i+3,i-1}, wherein he discussed quantum corrections in a curved space-time, Seeley's works \cite{i+6,j+4,i+8} for the study of boundary value problems, as well as Gilkey's one \cite{j+7} on spectral geometry. These works moved the method to a qualitatively higher level. Examples of using the approach in theoretical physics are the Casimir effect \cite{6,r}, research of anomalies of chiral gauge theories \cite{r+1}, calculation of the first correction for a black hole entropy \cite{r+5}, renormalization the Yang--Mills quantum theory in the background field formalism in two and three loops \cite{r+6,r+7}, as well as many others. \\

\noindent The main interest of mathematicians to the heat kernel  appeared due to the index theorem, which was first proved by Atiyah and Singer in their work \cite{j} in 1963. With the release of Patodi's paper \cite{j+5}, in which some reduction were shown, the connection between the index theorem and the asymptotic decomposition of the heat kernel began to be actively used in the proofs \cite{j+6}, which has a tendency to this day. \\

\noindent Speaking about the construction of the heat kernel, it is important to distinguish manifolds without boundary from one with boundary. In the second case, the problem is supplemented with boundary conditions that the heat kernel must satisfy. For the first time this statement appeared in the McKean and Singer work \cite{t}. Further, in the McAvity and Osborn articles \cite{y+4,y+5,y+6} cases with different boundary conditions were considered including generalized ones. \\

\noindent As examples of applications in mathematics and mathematical physics, one can cite the study of the spectral density of the Klein-Gordon operator \cite{j+7}, the calculation of the late time asymptotics of the heat kernel \cite{y+3}, researching of the problems with singular potential \cite{z+8}, domain walls \cite{j+8} or on the ball \cite{m}. Many other applications can be found in the fairly detailed review \cite{t+5}. \\

\noindent Due to the emergence of functional integration \cite{z} and the concept of Feynman diagrams \cite{z+2} in theoretical physics, many mathematical objects have acquired a clear physical meaning. In particular, this approach found its application in the theory of the heat kernel \cite{z+3,z+4,z+5}. As known, the work with the path integral requires the determination of the operator determinant, however, in most cases, such values diverge. One way to bring clarity was proposed in \cite{i+4,i+5} and is based on regularization by analytic continuation of the Riemann zeta function. \\

\noindent An explicit formula for the heat kernel can be obtained only in some special cases \cite{t+14}, therefore the calculation of the coefficients of the asymptotic series and their diagonal elements became a separate important task. In addition to the long-known results for the first three coefficients \cite{j+7, t}, it is important to recall the calculation of the fourth \cite{t+1, t+2} and the fifth \cite {t+4} coefficients in the most general formulation on manifolds without boundary. In this case, the approaches can be divided into two parts: non-recursive \cite{t+7,t+10} and recursive. A description of the last one can be found in \cite{t+11,t+12,t+13}, where the technique is based on the expansion of the operator exponent, or in \cite{t+8}, which is using the decomposition of the unit, and in \cite{t+9, t+10}, which is based on the formula for differentiating the Wilson line along a geodesic line. When one add a boundary and boundary conditions, the task acquires a more intricate view, so the counting method becomes more complicated. Some examples of the first terms calculations for the asymptotics can be found in \cite{y+7,y+8,y+10}. \\

\noindent From a brief historical background, it follows that in addition to an exact mathematical description, there is a more physical approach, which  consists both in the formal construction of asymptotics and in the use of functional integration. \\

\noindent  This paper is devoted to a detailed description of a model problem in a domain without boundary with the Euclidean metric and arbitrary smooth components of the gauge connection and potential. Such a model problem can be further used to work with more complex cases (curved metric, manifold with a boundary). The purpose of the article is to show the connection of two approaches, the first of which is based on the asymptotic expansion and is mainly used to find the diagonal elements $a_n(x,x)$ (see (\ref{x1})), and the second one uses the path intergation that clarifies the physical meaning. The paper includes a new non-recursive formula and proofs. \\

\noindent The work contains two main parts. Firstly, in the Section \ref{3}, we study the formulae for the differentiation of the ordered exponentials, the covariant expansion and the Fock-Schwinger gauge. Thereafter the simplest way is offered to obtain $a_n(x,x)$, based on combining previously known methods, as well as the derivation of the non-recursive formula in the general formulation. In view of this, readers who are interested only in calculations, we immediately redirect to Theorems \ref{pqs} and \ref{pq}. \\

\noindent Further, in the Section \ref{4}, we give the original derivation (without taking the limit) of the formula with the path integral (Theorem \ref{bg}) and the proof of equality of approaches at small times using the exponential formula (Theorem \ref{uy}). The paper also discusses the dependence on gauge, the presence of a curved metric, and examples of calculations.

\section{Problem statement}
\label{2}
Let $d$  be the dimension of a space, and $\Omega\subset\mathbb{R}^d$ be a smooth convex domain without a boundary of dimension  $\dim\Omega=d$.
By the Greek letters $\mu,\nu,\rho,\ldots$ we denote elements from $\{1,\ldots,d\}$. 
Let also a compact group $G$ and its Lie algebra $\mathfrak{g} $ be given, then we can introduce a gauge connection whose elements $ B_{\mu}\in \mathfrak{g} $ are smooth functions on $\Omega$. In this case, the covariant derivative is $D_{x^{\mu}} = I_G \partial_{x^{\mu}} + B_{\mu}(x)$, where $ I_G \in G $ is the unit in the group. Next, we define a potential $ v(x) $, that is square matrix of  order $m$, smooth on $ \Omega $, where $ m \in \mathbb {N} $. Then the Laplace operator has the form

\begin{equation}
A(x)=-ID_{x_{\mu}}D_{x^{\mu}}-v(x),
\end{equation}
where $ I $ is the unit matrix of order $ m $, and $ x \in\Omega $. Further, $ I_G $ and $ I $ will be omitted.
From the general theory it is well known that the problem for finding the heat kernel is the following
\begin{equation}
\label{po}
\begin{cases}
\text{$\left(\frac{\partial}{\partial\tau}+A(x)\right)K(x,y;\tau)=0;$}\\
\text{$K(x,y;0)=\delta(x-y),\,\,x,y\in\Omega,\,\,\tau\geqslant0.$}
\end{cases}
\end{equation}
In this case, the ansatz is chosen as a series
\begin{equation}
\label{x1}
K(x,y;\tau)=(4\pi\tau)^{-\frac{d}{2}}
e^{-\frac{(x-y)^2}{4\tau}}\sum\limits_{n=0}^{\infty}\tau^n\hat{a}_n(x,y),
\end{equation}
where in the right part in front of the sum the fundamental solution of the operator $-\partial_{x_{\mu}}\partial_{x^{\mu}}$ is allocated, which is a $\delta$ - sequence as $\tau\rightarrow+0$. Substituting the ansatz into the problem, one can find a system of recurrence equations: 
\begin{equation}\label{e3}
(x-y)^{\mu}D_{x^{\mu}}\hat{a}_0(x,y)=0,\,\,\hat{a}_0(x,x)=1;
\end{equation}
\begin{equation}\label{e4}
(n+(x-y)^{\mu}D_{x^{\mu}})\hat{a}_n(x,y)=-A(x)\hat{a}_{n-1}(x,y),\,\,n>0.
\end{equation}
It is important to note that in a formal construction, we can choose an arbitrary matrices with a size of $n\times n$, where $n\in\mathbb {N}$, as elements of  connection. This will have no effect on formal constructions, but the meaning of the covariant derivative and the choice of the gauge condition may be lost (see \cite{f,f+1}). You can also assume that $\Omega=\mathbb{R}^d$.

\section{First approach}
\label{3}
To find the coefficients of the asymptotic expansion (\ref{x1}), we propose to use the recurrence relations (\ref{e3}) and (\ref{e4}). As noted in the introduction, there are many variations of building techniques. This section will show the simplest way, which, moreover, provides a non-recursive answer for diagonal contributions $\hat{a}_n(x,x),\,\,n>0,\,\,x\in\Omega$.

\subsection{Path-ordered exponential}
\label{31}
When working with the heat kernel, the ordered exponential (see \cite{f+2}) is crucial and can appear in a wide of variations (see formulae (\ref{bv}), (\ref{bv1}), (\ref{e2}), or (\ref{bv2})). As a rule, it is the function not only of the integration limits, but also of a contour $\gamma$  along which the ordering is performed. Let us consider the simplest case when a straight line is chosen as the parameterization.
Let $x,y\in\Omega$ and $z^{\mu}(s)=(1-s)y^{\mu}+sx^{\mu}$, where $s\in[0,1]$, then
\begin{equation}
\label{bv}
\Phi(x,y)=1+\sum\limits_{k=1}^{\infty}(-1)^k\int_0^1dz^{\mu_1}(s_1)\ldots\int_0^{s_{k-1}}dz^{\mu_k}(s_k)\,B_{\mu_1}(z(s_1))\ldots B_{\mu_k}(z(s_k)).
\end{equation}
\noindent This function has a number of important properties. Let
$x,y,z\in\Omega$ and $\exists\,s\in\mathbb{R}:z^{\mu}(s)=(1-s)y^{\mu}+sx^{\mu}$, then
\begin{equation}
\Phi(x,x)=1,\,\,\Phi(x,y)=\Phi^{-1}(y,x),\,\,\Phi(x,z)\Phi(z,y)=\Phi(x,y).
\end{equation}
Taking into account the latter properties, one can write out the solution of the recurrence system (see \cite{t+10}).
\begin{lemma}\label{e7}
Let $x,y\in\Omega$ and $z^{\mu}(s)=(1-s)y^{\mu}+sx^{\mu}$, where $s\in[0,1]$, then
\begin{equation}
\hat{a}_0(x,y)=\Phi(x,y),\,\,
\hat{a}_n(x,y)=-\int_0^1ds\,s^{n-1}\Phi(x,z(s))A(z(s))\hat{a}_{n-1}(z(s),y),\,\,n>0.
\end{equation}
\end{lemma}

\subsection{Covariant expansion}
\label{32}
The covariant expansion is an analogue of the Taylor series for the noncommutative case and is realized by analyzing $\Phi(y,x)f(x)\Phi(x,y)$ instead of $f(x)$. Due to the cumbersome formulae, it is convenient to introduce additional notation.
Let $f\in C(\Omega,\mathfrak{g})$ and $x\in\Omega$, then
\begin{equation*}
\overrightarrow{D}_{x^{\mu}} f(x)=\partial_{x^{\mu}}f(x)+B_{\mu}(x)f(x),\,\,
f(x)\overleftarrow{D}_{x^{\mu}}=\partial_{x^{\mu}}f(x)-f(x)B_{\mu}(x).
\end{equation*}
To prove the basic Lemma on expansion into the covariant series, the first derivatives of $\Phi(x, y)$ are required in a special form.
\begin{lemma}\label{r1} Let $x,y\in\Omega$, $s\in[0,1]$, $z^{\mu}(s)=(1-s)y^{\mu}+sx^{\mu}$, $w^{\mu}(s)=(1-s)x^{\mu}+sy^{\mu}$, and $F_{\mu\nu}(x)=[D_{x^{\mu}}, D_{x^{\nu}}]$, then
\begin{equation}\label{eq1}
\overrightarrow{D}_{x^{\mu}}\Phi(x,y)=\int_0^1ds\,s(x-y)^{\nu}\Phi(x,z(s))F_{\nu\mu}(z(s))\Phi(z(s),y),
\end{equation}
\begin{equation}\label{eq2}
\Phi(x,y)\overleftarrow{D}_{y^{\mu}}=-\int_0^1ds\,s(y-x)^{\nu}\Phi(x,w(s))F_{\nu\mu}(w(s))\Phi(w(s),y).
\end{equation}
\end{lemma}
\noindent The formula (\ref{eq2}) follows from the equality $\partial_{y^{\mu}}\Phi(x,y)\Phi(y,x)=0$ and the formula (\ref{eq1}), the proof of which (see \cite{f+3,f+4}) can be obtained by direct differentiation of the equality (\ref{bv}) and further integration by parts.
\begin{lemma}\label{r2}
Let $x,y\in\Omega$ and $f\in\mathbb{C}^\infty(\Omega,\mathfrak{g})$, then
\begin{equation}\label{eq3}
\Phi(y,x)f(x)\Phi(x,y)=\sum_{n=0}^\infty\frac{(x-y)^{\mu_1\ldots\mu_n}}{n!}\nabla_{y^{\mu_1}}\ldots\nabla_{y^{\mu_n}}f(y),
\end{equation}
where $\nabla_{x^{\mu}}\,\cdot=\partial_{x^{\mu}}\cdot+[B_{\mu}(x), \cdot \ ]$ and $(x-y)^{\mu_1\ldots\mu_n}=(x-y)^{\mu_1}\ldots(x-y)^{\mu_n}$.
\end{lemma}
\noindent\textbf{Proof:} Let $n\in\mathbb{N}$.
Let us consider a derivative of order $n$ and use the formulae (\ref{eq1}) and (\ref{eq2}), then the following expansion is performed
\begin{equation*}
\partial_{x^{\mu_1}}\ldots\partial_{x^{\mu_n}}\Phi(y,x)f(x)\Phi(x,y)=
\Phi(y,x)\left(\nabla_{x^{\mu_1}}\ldots\nabla_{x^{\mu_{n}}}f(x)\right)\Phi(x,y)+\ldots,
\end{equation*}
where dots denote terms containing $F_{\mu\nu}$ and its derivatives.
The expansion shows that after multiplying by $(x-y)^{\mu_1...\mu_n}$ all terms except the first one are zero because of the contraction 
$(x-y)^{\mu_i}(x-y)^{\mu_j}F_{\mu_i\mu_j}=0$. Therefore, 
\begin{equation*}
\frac{(x-y)^{\mu_1\ldots\mu_n}}{n!}\left(\partial_{x^{\mu_1}}\ldots\partial_{x^{\mu_n}}\Phi(y,x)f(x)\Phi(x,y)\bigg|_{x=y}\right)=\frac{(x-y)^{\mu_1\ldots\mu_n}}{n!}\nabla_{y^{\mu_1}}\ldots\nabla_{y^{\mu_n}}f(y),
\end{equation*}
from what the statement of the Lemma follows. $\blacksquare$

\subsection{Fock-Schwinger gauge}
\label{33}
\noindent For certainty, we introduce two functions of two variables, which are from $\mathbb{R}^d\times\Omega$:
\begin{equation}
f_{\mu}(x-y,y)=\sum_{k=1}^\infty\frac{(x-y)^{\mu_1\ldots\mu_k}}{(k+1)!}\left(\nabla_{y^{\mu_1}}\ldots\nabla_{y^{\mu_{k-1}}}F_{\mu_k\mu}(y)\right);
\end{equation}
\begin{equation}\label{e1}
f^{FS}_{\mu}(x-y,y)=(x-y)^{\nu}\partial_{x^{\nu}}f_{\mu}(x-y,y).
\end{equation}
From the definitions it is clear that both functions are mappings from $\mathbb{R}^d\times\Omega$ to $\mathfrak{g}$ for fixed $\mu$ and satisfy the Fock-Schwinger gauge condition, that is 
$$  (x-y)^{\mu}f_{\mu}(x-y,y)=0,\,\,(x-y)^{\mu}f^{FS}_{\mu}(x-y,y)=0.$$ 
Sometimes such gauge condition is called the radial \cite{f+5} or relativistic Poincaré \cite{f+6,f+7} one.
\begin{theorem}
\label{we}
Let $x,y\in\Omega$ and $\mu\in\{1,\ldots,d\}$, then we have the equalities
\begin{equation}
\label{r3}
\overrightarrow{D}_{x^{\mu}}\Phi(x,y)=\Phi(x,y)f^{FS}_{\mu}(x-y,y),
\end{equation}
\begin{equation}\label{r4}
\Phi(x,y)\overleftarrow{D}_{y^{\mu}}=-f^{FS}_{\mu}(y-x,x)\Phi(x,y),
\end{equation}
\begin{equation}\label{r5}
\Phi(x,y)\overleftarrow{D}_{y^{\mu}}=\Phi(x,y)f_{\mu}(x-y,y),
\end{equation}
\begin{equation}\label{r6}
\overrightarrow{D}_{x^{\mu}}\Phi(x,y)=-f_{\mu}(y-x,x)\Phi(x,y).
\end{equation}
\end{theorem}
\noindent\textbf{Proof}:
Let us apply Lemma \ref{r2} to the function $F_{\mu\nu}$ at  the point $z^{\mu}(s)=(1-s)y^{\mu}+sx^{\mu}$:
\begin{equation*}
\Phi(y,z)F_{\nu\mu}(z)\Phi(z,y)=\sum_{k=0}^\infty\frac{(z-y)^{\mu_1\ldots\mu_k}}{k!}\nabla_{y^{\mu_1}}\ldots\nabla_{y^{\mu_k}}F_{\nu\mu}(y).
\end{equation*}
Then we can rewrite the integral from the formula (\ref{eq1}) as
\begin{equation*}
\int_0^1ds\,s(x-y)^{\nu}\Phi(x,z)F_{\nu\mu}(z)\Phi(z,y)
=\Phi(x,y)\sum_{k=1}^{\infty}\frac{(x-y)^{\mu_1\ldots\mu_k}}{(k+1)!}k\nabla_{y^{\mu_1}}\ldots\nabla_{y^{\mu_{k-1}}}F_{\mu_k\mu}(y),
\end{equation*}
from which the equality (\ref{r3}) follows.  Further, choosing the parameterization in the following form $w^{\mu}(s)=(1-s)x^{\mu}+sy^{\mu}$,
we obtain the relation (\ref{r4}). Analyzing in a similar way the second integral from Lemma \ref{r1}, one can verify the equalities
(\ref{r5}) and (\ref{r6}). $\blacksquare$
\begin{corollary}Let $x,y\in\Omega$ and $\mu\in\{1,\ldots,d\}$, then we have
\begin{equation}
\Phi(x,y)f_{\mu}(x-y,y)=-f^{FS}_{\mu}(y-x,x)\Phi(x,y),
\end{equation}
\begin{equation}
\Phi(x,y)\left(f_{\mu}(x-y,y)+f^{FS}_{\mu}(x-y,y)\right)=-\left(f_{\mu}(y-x,x)+f^{FS}_{\mu}(y-x,x)\right)\Phi(x,y).
\end{equation}
\end{corollary}
\begin{corollary}Let $x,y\in\Omega$ and $\mu\in\{1,\ldots,d\}$, then we have
\begin{equation}
B_{\mu}(x)=\Phi(x,y)\left[
(x-y)^{\nu}\partial_{x^{\nu}}\Phi(y,x)\partial_{y^{\mu}}\Phi(x,y)\right]
\Phi(y,x)-\left[\partial_{x^{\mu}}\Phi(x,y)\right]\Phi(y,x).
\end{equation}
\end{corollary}
\noindent The first consequence follows from the combination of the formulae (\ref{r3}) - (\ref{r6}), the second one follows from the equations (\ref{r3}) and (\ref{r5}), and the definition (\ref{e1}). Let us rewrite the relation (\ref{r3}) in the form
\begin{equation}
f^{FS}_{\mu}(x-y,y)=
\Phi^{-1}(x,y)\partial_{x^{\mu}}\Phi(x,y)+
\Phi^{-1}(x,y)B_{\mu}(x)\Phi(x,y).
\end{equation}
It shows that $ f^{FS}_{\mu} (x-y, y) $ is the gauge transformation of the connection component $B_{\mu}\in\mathfrak{g}$ by using $\Phi(x, y)\in\mathcal{G}$, where $\mathcal{G}$ is a group of gauge transformations.

\subsection{Solution of the recurrence system}
\label{34}
Let us return to the Seeley--DeWitt coefficients. Using the substitution $\hat{a}(x,y)=\Phi(x,y)a(x,y)$ we move in the Fock-Schwinger gauge. Applying Lemmas \ref{e7} and \ref{r2}, and Theorem \ref{we}, the recurrence system of the equations (\ref{e3}) - (\ref{e4}) can be rewritten in the following form
\begin{equation}\label{e12}
a_0(x,y)=1,\,\,
(n+(x-y)^{\mu}\partial_{x^{\mu}})a_n(x,y)=-A(x-y,y)a_{n-1}(x,y),\,\,n>0,
\end{equation}
where
\begin{equation}
\label{w5}
A(x-y,y)=\Phi(y,x)A(x)\Phi(x,y)=-(\partial+f^{FS}(x-y,y))_{\mu}(\partial+f^{FS}(x-y,y))^{\mu}-V(x-y,y)
\end{equation}
and the relation holds
\begin{equation}
\label{w6}
V(x-y,y)=\Phi(y,x)v(x)\Phi(x,y)=\sum_{k=0}^\infty\frac{(x-y)^{\mu_1\ldots\mu_k}}{k!}\nabla_{y^{\mu_1}}\ldots\nabla_{y^{\mu_k}}v(y).
\end{equation}
\begin{lemma} Let $x,y\in\Omega$, then the solution of the system (\ref{e12}) is
\begin{equation}
\label{w1}
a_n(x,y)=(-1)^n\int_0^1ds_n\ldots\int_0^{s_2}ds_1\,A(s_n(x-y),y)\ldots A(s_1(x-y),y)\cdot1,\,\,n\geqslant0,
\end{equation}
and 
\begin{equation}
\label{bv1}
\sum\limits_{n=0}^{\infty}\tau^na_n(x,y)=P_se^{-\tau\int_0^1ds\,A(s(x-y),y)}\cdot1,\,\,
\tau\rightarrow+0.
\end{equation}
\end{lemma}
\noindent For the proof it is necessary to use the induction and the definition of the ordered exponential.

\subsection{The method of calculating}
\label{35}
\noindent\textbf{Notations:} 
We introduce the objects by which operators will be numbered. Using the italicized Greek symbol with the index $\mbox{\boldmath$\mu_i$}$ we denote the pair $(\mu_i,i)$, where $\mu_i\in\{1,\ldots,d\}$ and $i\in\mathbb{N}$. Thus the equation $\mbox{\boldmath$\mu_i$}=\mbox{\boldmath$\mu_j$}$ is equivalent to the relations $\mu_i=\mu_j$ and $i=j$. Also, if double repeated indices  $\mu_i\mu_i$  appear the summation from $1$ to $d$ is implied. Further, using the round brackets we denote the symmetrization. For example, a tensor with index $(\mu_{i_k}\ldots\mu_{i_1})$ is equal to the  sum of tensors over all possible index permutations devided by $k!$. In this case, the allocation of a group of indices by the symbol $|$ means that they are not symmetrized. Also, if $\sigma=\{\sigma(p),\ldots,\sigma(1)\}$ is a set of numbers, then by definition we denote $\mu_{\sigma}=\mu_{\sigma(p)}\ldots\mu_{\sigma(1)}$.\\

\noindent\textbf{Operator definitions:} Let us introduce a set of operators and define the necessary properties among them. In total, four types are required: $\mbox{\boldmath$A^{\mu_j}$}$ is a differentiation operator, $\mbox{\boldmath$S$}^k$ is an integration operator, where $k>0$, an operator of multiplication $ \mbox{\boldmath$\Omega^{\mu_{i_k}\ldots\mu_{i_1}}$}$, where $i_k\geqslant\ldots\geqslant i_2\geqslant i_1>0,\,\,k>0$, as well as the operator $\mbox{\boldmath$\Upsilon$}$, which maps the tensor to the multiplication operator. So, first we define the initial element $\mathbb{1}$ 
by relation
\begin{equation}
\label{w4}
\mbox{\boldmath$\Upsilon$}\mathbb{1}=1,\,\,\mbox{\boldmath$A^{\mu_j}$}\mathbb{1}=0,
\end{equation}
and then we give sufficient to work relations:
\begin{equation}
\label{w2}
\mbox{\boldmath$\Upsilon$}\mbox{\boldmath$\Omega^{\mu_{i_k}\ldots\mu_{i_1}}$}=
\mbox{\boldmath$\tilde{\Omega}^{\mu_{i_k}\ldots\mu_{i_1}}$}\mbox{\boldmath$\Upsilon$},\,\,
\mbox{\boldmath$\Upsilon$}\mbox{\boldmath$S$}^k=\frac{1}{k}\mbox{\boldmath$\Upsilon$},
\end{equation}
\begin{equation}
\label{w3}
\mbox{\boldmath$A^{\mu_j}$}\mbox{\boldmath$\Omega^{\mu_{i_k}\ldots\mu_{i_1}}$}=
\mbox{\boldmath$\Omega^{\mu_j\mu_{i_k}\ldots\mu_{i_1}}$}+
\mbox{\boldmath$\Omega^{\mu_{i_k}\ldots\mu_{i_1}}$}
\mbox{\boldmath$A^{\mu_j}$},\,\,
\mbox{\boldmath$A^{\mu_j}$}\mbox{\boldmath$S$}^k=\mbox{\boldmath$S$}^{k+1}
\mbox{\boldmath$A^{\mu_j}$},
\end{equation}
where $k>0$, $j\geqslant i_k\geqslant\ldots\geqslant i_2\geqslant i_1>0$, and tensors $\mbox{\boldmath$\tilde{\Omega}^{\mu_{i_k}\ldots\mu_{i_1}}$}$ are determined by:\\
if $i_2>i_1$, then for $k>1$ we have
\begin{equation}
\label{tr}
\mbox{\boldmath$\tilde{\Omega}^{\mu_{i_k}\ldots\mu_{i_1}}$}=\frac{2(k-1)}{k}\nabla_{(\mu_{i_k}}\ldots\nabla_{\mu_{i_3}}F_{\mu_{i_2})\mu_{i_1}}(x);
\end{equation}
if $i_2=i_1$, then
\begin{equation}
\mbox{\boldmath$\tilde{\Omega}^{\mu_{i_2}\mu_{i_1}}$}=v(x);
\end{equation}
\begin{equation}
\mbox{\boldmath$\tilde{\Omega}^{\mu_{i_3}\mu_{i_2}\mu_{i_1}}$}=\frac{2}{3}\nabla_{(\mu_{i_3}}F_{\mu_{i_1})\mu_{i_1}}(x)+\nabla_{\mu_{i_3}}v(x);
\end{equation}
and for $k>3$ we have
\begin{multline}
\label{q1}
\mbox{\boldmath$\tilde{\Omega}^{\mu_{i_k}
\ldots\mu_{i_1}}$}=
\frac{k-1}{k}\nabla_{(\mu_{i_k}}\ldots\nabla_{\mu_{i_3}}F_{\mu_{i_1})\mu_{i_1}}(x)
+\nabla_{(\mu_{i_k}}\ldots\nabla_{\mu_{i_3})}v(x)+\\+
\sum\limits_{n=0}^{k-4}\frac{(n+1)(k-n-3)(k-2)!}{(n+2)!(k-n-2)!}
\nabla_{(\mu_{i_{k}}}\ldots\nabla_{\mu_{i_{n+5}}}F_{\mu_{i_{n+4}}|\mu_{i_1}|}(x)
\nabla_{\mu_{i_{n+3}}}\ldots\nabla_{\mu_{i_4}}F_{\mu_{i_3})\mu_{i_1}}(x).
\end{multline}
If there is only one index, the function is identically zero $\mbox{\boldmath$\tilde{\Omega}^{\mu_{i}}$}=0$. It is also convenient to add notations by $\mbox{\boldmath$\tilde{\Omega}^{\mu_{\sigma}}$}=1$, if $\mbox{\boldmath$\sigma$}=\emptyset$, this allows us to write the answer in a more compact form.
\begin{lemma}Let $x\in\Omega$ and $n>0$, then, taking into account the definitions (\ref{w4})-(\ref{q1}), we have
\begin{equation}
\label{w7}
a_{n}(x,x)=\mbox{\boldmath$\Upsilon$}\prod\limits_{k=1}^n\mbox{\boldmath$S$}^k\left(
\mbox{\boldmath$A^{\mu_{2k}}$}\mbox{\boldmath$A^{\mu_{2k}}$}+
\mbox{\boldmath$\Omega^{\mu_{2k}}$}
\mbox{\boldmath$A^{\mu_{2k}}$}+
\mbox{\boldmath$\Omega^{\mu_{2k}\mu_{2k}}$}
\right)\mathbb{1}.
\end{equation}
\end{lemma}
\noindent\textbf{Proof:} The statement follows from substitution of the definitions (\ref{w5}) and (\ref{w6}) into the formula (\ref{w1}) and considering the Laplace operator in the form
\begin{multline}
-A(x-y,y)=\partial_{x_{\mu}}\partial_{x^{\mu}}+2f^{FS}_{\mu}(x-y,y)\partial_{x_{\mu}}+\\+\partial_{x_{\mu}}f^{FS}_{\mu}(x-y,y)+
f^{FS}_{\mu}(x-y,y)f^{FS\mu}(x-y,y)+V(x-y,y).
\end{multline}
The property (\ref{w3}) of the integration operator takes into account the relationship between the degree of parameterization parameter and the number of derivatives acting on the integrand in the formula (\ref{w1}). $\blacksquare$\\
\noindent In the formula (\ref{w7}) the symbol $\mbox{\boldmath$\mu_{2k}$}$  has even index. This is done specifically for the convenience of the following theorem and does not affect the answer, since $\mbox{\boldmath$\mu_{2k}$}$ are dumb and the ordering, which is important in the definitions (\ref{tr}) - (\ref{q1}), is not violated.
\begin{theorem}
\label{pqs}
Let $I_{2n}=\{2n,\ldots,1\}$ and  $\mbox{\boldmath$\sigma$}_n$ is a set of all sets $\{\sigma_{i}\}_{i=1}^{n}$,
such that $\sigma_i=\{\sigma_i(p_i),\ldots,\sigma_i(1)\}$, where $p_i=\sharp\sigma_i$,
satisfying the following conditions:
\begin{enumerate}
\item$\sigma_{i}(k)>\sigma_{i}(j)$ for $p_i\geqslant k>j\geqslant1;$
\item$\cup_{j=1}^n\sigma_{j}=I_{2n};$
\item$\sigma_{i}\subset\{2n,\ldots,2i-1\}\setminus(\cup_{j=i+1}^{n}\sigma_{j});$
\item either $\{2i-1\}\subset\sigma_{i}$ or $\sigma_{i}=\emptyset,$
\end{enumerate}
where $i\in\{1,\ldots,n\}$. Then the diagonal part of $\hat{a}_n(x,y)$ from the formula (\ref{x1}) is equal to
\begin{equation}
\label{w8}
a_{n}(x,x)=\left.
\sum\limits_{\mbox{\boldmath$\sigma$}_n}
\frac{\mbox{\boldmath$\tilde{\Omega}^{\mu_{\sigma_{n}}}$}}{\sum_{i=1}^np_i-n}\ldots
\frac{\mbox{\boldmath$\tilde{\Omega}^{\mu_{\sigma_{2}}}$}}{p_1+p_2-2}
\frac{\mbox{\boldmath$\tilde{\Omega}^{\mu_{\sigma_{1}}}$}}{p_1-1}\right|_{\mu_{2j-1}=\mu_{2j},\forall j}.
\end{equation}
\end{theorem}
\noindent\textbf{Proof:} 
In oder to show the equality of the formulae (\ref{w7}) and (\ref{w8}), it is convenient to prove a more general relation. We show that for $n>0$ we have the equality
\begin{multline}
\label{re}
\prod\limits_{i=1}^n\mbox{\boldmath$S$}^i\left(
\mbox{\boldmath$A^{\mu_{2i}}$}\mbox{\boldmath$A^{\mu_{2i}}$}+
\mbox{\boldmath$\Omega^{\mu_{2i}}$}
\mbox{\boldmath$A^{\mu_{2i}}$}+
\mbox{\boldmath$\Omega^{\mu_{2i}\mu_{2i}}$}
\right)\mathbb{1}=\\=
\left.
\sum\limits_{\mbox{\boldmath$\sigma$}_n}
\mbox{\boldmath$S$}^{\sum_{i=1}^np_i-n}
\mbox{\boldmath$\Omega^{\mu_{\sigma_{n}}}$}\ldots
\mbox{\boldmath$S$}^{p_1+p_2-2}
\mbox{\boldmath$\Omega^{\mu_{\sigma_{2}}}$}
\mbox{\boldmath$S$}^{p_1-1}
\mbox{\boldmath$\Omega^{\sigma_{1}}$}\mathbb{1}
\right|_{\mu_{2j-1}=\mu_{2j},\forall j},
\end{multline}
then, applying the operator $\mbox{\boldmath$\Upsilon$}$, we obtain the statement of the Lemma.
The proof is convenient to produce by induction. Indeed, for $n=1$ the formula is obvious. Assume that the equality is true for $n=k-1$ and prove it for $n=k$. To do this, one can  apply the operator 
$\mbox{\boldmath$S$}^k(\mbox{\boldmath$A^{\mu_{2k}}$}\mbox{\boldmath$A^{\mu_{2k}}$}+
\mbox{\boldmath$\Omega^{\mu_{2k}}$}\mbox{\boldmath$A^{\mu_{2k}}$}+\mbox{\boldmath$\Omega^{\mu_{2k}\mu_{2k}}$})$.
Identification mark $\mbox{\boldmath$\mu_{2j-1}$}=\mbox{\boldmath$\mu_{2j}$}$ on the right side (\ref{re}) works for $j\in\{1,\ldots,k-1\}$,  so we can add $\mbox{\boldmath$\mu_{2k-1}$}=\mbox{\boldmath$\mu_{2k}$}$ and put the operator under the identification mark in the form 
$\mbox{\boldmath$S$}^k(\mbox{\boldmath$A^{\mu_{2k}}$}\mbox{\boldmath$A^{\mu_{2k-1}}$}+
\mbox{\boldmath$\Omega^{\mu_{2k-1}}$}\mbox{\boldmath$A^{\mu_{2k}}$}+\mbox{\boldmath$\Omega^{\mu_{2k}\mu_{2k-1}}$})$.
Let us consider separately each term in order to understand the transition $\mbox{\boldmath$\sigma$}_{k-1}\rightarrow\mbox{\boldmath$\sigma$}_{k}$:\\
\noindent1) The term with $\mbox{\boldmath$S$}^k\mbox{\boldmath$\Omega^{\mu_{2k}\mu_{2k-1}}$}$ is described by permutations $\{\sigma_{i}\}_{i=1}^{k}$, where $\sigma_{k}=\{2k,2k-1\}$ and $\{\sigma_{i}\}_{i=1}^{k-1}\in\mbox{\boldmath$\sigma$}_{k-1}$. Considering that $p_k=2$ and $\sum_{i=1}^{k-1}p_i=2k-2$, we can write $\mbox{\boldmath$S$}^k=\mbox{\boldmath$S$}^{\sum_{i=1}^{k}p_i-k}$;\\
\noindent2) The term with $\mbox{\boldmath$S$}^k\mbox{\boldmath$\Omega^{\mu_{2k-1}}$}\mbox{\boldmath$A^{\mu_{2k}}$}$
corresponds to the set in which $\sigma_{k}=\{2k-1\}$ and $\{\sigma_{i}\}_{i=1}^{k-1}$ is constructed using a set from $\mbox{\boldmath$\sigma$}_{k-1}$ by combining one of its elements with $\{2k\}$. Wherein $p_k=1$ and $\sum_{i=1}^{k-1}p_i=2k-1$;\\
\noindent3) The term with $\mbox{\boldmath$S$}^k\mbox{\boldmath$A^{\mu_{2k}}$}\mbox{\boldmath$A^{\mu_{2k-1}}$}$ responds to $\sigma_{k}=\emptyset$ and $\{\sigma_{i}\}_{i=1}^{k-1}$, which is constructed by adding $2k$ and $2k-1$ in the elements of the set from $\mbox{\boldmath$\sigma$}_{k-1}$. Wherein $p_k=0$ and $\sum_{i=1}^{k-1}p_i=2k$.
$\blacksquare$

\subsection{Examples}
\label{36}
As a small exercise, we give a calculation of the diagonal parts of the coefficients $a_n(x,y)$ of the heat kernel for $n=1,2,3$.\\

\noindent $n=1$: in this case there is only one term, and it is equal to: $a_1(x,x)=\mbox{\boldmath$\tilde{\Omega}^{\mu_1\mu_1}$}=v(x)$.\\

\noindent $n=2$: then the formula is
\begin{equation}
a_2(x,x)=\left.\sum\limits_{\mbox{\boldmath$\sigma$}_2}
\frac{\mbox{\boldmath$\tilde{\Omega}^{\mu_{\sigma_{2}}}$}}{(p_1+p_2-2)}
\frac{\mbox{\boldmath$\tilde{\Omega}^{\mu_{\sigma_{1}}}$}}{(p_1-1)}\right|_{\mu_{2j-1}=\mu_{2j},\forall j}.
\end{equation}
Let us consider all possible subsets $\{\sigma_{i}\}^2_{i=1}$ from the set $\mbox{\boldmath$\sigma$}_{2}$:
$$\{4,3\}\{2,1\},\,\,\{3\}\{4,2,1\},\,\,\emptyset\,\{4,3,2,1\}.$$
Since $\mbox{\boldmath$\tilde{\Omega}^{\mu_3}$}=0$ and $\nabla_{(\mu_4}\nabla_{\mu_4}F_{\mu_2)\mu_2}=0$, we get
\begin{equation}
\frac{1}{2}
\mbox{\boldmath$\tilde{\Omega}^{\mu_4\mu_4}$}
\mbox{\boldmath$\tilde{\Omega}^{\mu_2\mu_2}$}+\frac{1}{6}
\mbox{\boldmath$\tilde{\Omega}^{\mu_4\mu_4\mu_2\mu_2}$}=
\frac{1}{2}v^2(x)+\frac{1}{12}F_{\mu_4\mu_2}(x)F_{\mu_4\mu_2}(x)+\frac{1}{6}\nabla_{\mu_4}\nabla_{\mu_4}v(x).
\end{equation}
\noindent $n=3$: the general formula for the coefficient is
\begin{equation}
a_3(x,x)=\left.\sum\limits_{
\mbox{\boldmath$\sigma$}_3}
\frac{
\mbox{\boldmath$\tilde{\Omega}^{\mu_{\sigma_{3}}}$}}{(p_1+p_2+p_3-3)}\frac{
\mbox{\boldmath$\tilde{\Omega}^{\mu_{\sigma_{2}}}$}}{(p_1+p_2-2)}\frac{
\mbox{\boldmath$\tilde{\Omega}^{\mu_{\sigma_{1}}}$}}{(p_1-1)}\right|_{\mu_{2j-1}=\mu_{2j},\forall j}.
\end{equation}
Taking into account the relations $\mbox{\boldmath$\tilde{\Omega}^{\mu_1}$}=0 $ and $
\mbox{\boldmath$\tilde{\Omega}^{\mu_6\mu_4}$}\mbox{\boldmath$\tilde{\Omega}^{\mu_6\mu_4\mu_2\mu_2}$}=0$,
we write out all nonzero sets $\{\sigma_{i}\}^3_{i=1}$:
$$\emptyset\,\,\emptyset\,\{6,6,4,4,2,2\},\,\,\emptyset\,\{4,4\}\{6,6,2,2\},\,\,\emptyset\,\{6,6,4\}\{4,2,2\},\,\,\emptyset\,\{6,4,4\}\{6,2,2\},$$
$$\emptyset\,\{6,4,4\}\{6,2,2\},\,\,\emptyset\,\{6,6,4,4\}\{2,2\},\,\,\{6,6\}\,\emptyset\,\{4,4,2,2\},\,\,\{6,6\}\{4,4\}\{2,2\}.$$
Since $\nabla_{(\mu_6}\nabla_{\mu_6}\nabla_{\mu_4}\nabla_{\mu_4}F_{\mu_2)\mu_2}=0$ and
$\nabla_{\mu_6}\nabla_{\mu_4}F_{\mu_6\mu_4}=0$, as well as using the properties
\begin{equation}
\nabla_\mu\nabla_\nu f=\nabla_\nu\nabla_\mu f+F_{\mu\nu}f+fF_{\nu\mu},\,\,\nabla_\mu F_{\nu\rho}+\nabla_\nu F_{\rho\mu}+\nabla_{\rho}F_{\mu\nu}=0,
\end{equation}
we get the final result for the coefficient:
\begin{multline}
a_3(x,x)=\frac{1}{45}\nabla_{\mu_6}F_{\mu_2\mu_4}\nabla_{\mu_6}F_{\mu_2\mu_4}+\frac{1}{180}\nabla_{\mu_6}F_{\mu_2\mu_6}\nabla_{\mu_4}F_{\mu_2\mu_4}+\\+\frac{1}{60}\nabla_{\mu_6}\nabla_{\mu_6}F_{\mu_2\mu_4}F_{\mu_2\mu_4}-\frac{1}{30}F_{\mu_2\mu_4}F_{\mu_4\mu_6}F_{\mu_6\mu_2}+\frac{1}{60}F_{\mu_2\mu_4}\nabla_{\mu_6}\nabla_{\mu_6}F_{\mu_2\mu_4}+\\+\frac{1}{6}v^3+\frac{1}{12}\nabla_{\mu_6}\nabla_{\mu_6}vv+\frac{1}{12}\nabla_{\mu_6}v\nabla_{\mu_6}v+\frac{1}{12}v\nabla_{\mu_6}\nabla_{\mu_6}v+\\+\frac{1}{30}vF_{\mu_2\mu_4}F_{\mu_2\mu_4}+\frac{1}{60}F_{\mu_2\mu_4}vF_{\mu_2\mu_4}+\frac{1}{30}F_{\mu_2\mu_4}F_{\mu_2\mu_4}v-\\-\frac{1}{60}\nabla_{\mu_2}v\nabla_{\mu_4}F_{\mu_4\mu_2}+\frac{1}{60}\nabla_{\mu_4}F_{\mu_4\mu_2}\nabla_{\mu_2}v+\frac{1}{60}\nabla_{\mu_6}\nabla_{\mu_6}\nabla_{\mu_4}\nabla_{\mu_4}v.
\end{multline}

\subsection{General case}
\label{37}
Similarly, by repeating step-by-step of the proof of Theorem \ref{pqs}, one can formulate a more general result.
\begin{theorem}
\label{pq}
Let $x,y\in\Omega$, $c^{\mu}, b\in C^{\infty}(\Omega)$, where $\mu\in\{1,\ldots,d\}$, $I_{2n}=\{2n,\ldots,1\}$ and  $\mbox{\boldmath$\sigma$}_n$ is the set of all subsets $\{\sigma_{i}\}_{i=1}^{n}$, satisfying four conditions from Theorem \ref{pqs}.
Then the diagonal part for the solution of the system of differential equations
\begin{equation}
\label{fer}
(n+(x-y)^{\mu}\partial_{x^{\mu}})a_n(x,y)=(\partial_{x_{\mu}}\partial_{x^{\mu}}+c^{\mu}\partial_{x^{\mu}}+b)a_{n-1}(x,y),\,\,n>0,
\end{equation}
with $a_0(x,y)=1$ is the following function
\begin{equation}
a_{n}(x,x)=\left.
\sum\limits_{\mbox{\boldmath$\sigma$}_n}
\frac{\mbox{\boldmath$\tilde{\Omega}^{\mu_{\sigma_{n}}}$}}{\sum_{i=1}^np_i-n}\ldots
\frac{\mbox{\boldmath$\tilde{\Omega}^{\mu_{\sigma_{2}}}$}}{p_1+p_2-2}
\frac{\mbox{\boldmath$\tilde{\Omega}^{\mu_{\sigma_{1}}}$}}{p_1-1}\right|_{\mu_{2j}=\mu_{2j-1},\forall j},
\end{equation}
where $\mbox{\boldmath$\tilde{\Omega}^{\mu_{i}}$}=c_{\mu_{i}}(x)$, $\mbox{\boldmath$\tilde{\Omega}^{\mu_{\sigma}}$}=1$, if $\sigma=\emptyset$, and for
$i_k\geqslant\ldots\geqslant i_2\geqslant i_1>0$, $k>1$:\\
if $i_2>i_1$, then
$\mbox{\boldmath$\tilde{\Omega}^{\mu_{i_k}\ldots\mu_{i_1}}$}=\partial_{(\mu_{i_k}}\ldots\partial_{\mu_{i_2})}c_{\mu_{i_1}}(x)$;\\
\noindent if $i_2=i_1$, then
$\mbox{\boldmath$\tilde{\Omega}^{\mu_{i_k}
\ldots\mu_{i_1}}$}=\partial_{(\mu_{i_k}}\ldots\partial_{\mu_{i_3})}b(x).$
\end{theorem}
\noindent For interest, we give the answers for $n=1,2$:
\begin{equation}
a_1(x,x)=b(x),\,\,\,\,a_2(x,x)=\frac{1}{2}b^2(x)+\frac{1}{4}c_{\mu}\partial_{x_{\mu}}b(x)+\frac{1}{6}\partial_{x_{\mu}}\partial_{x^{\mu}}b(x).
\end{equation}

\section{Second approach}
\label{4}
The approach to which this part of the work is devoted is focused on compact recording and clarification of physical meaning. The path integral is in some way a Golden mean, since it retains the necessary mathematical properties to work with physical phenomena, although it still contains a lot of open questions. The transition from the previously studied asymptotic expansion (\ref{x1}) to the path integral representation is carried out in two stages. First, we prove that at small times the asymptotic series can be written as an exponential formula (\ref{e2}), and then, based on the basic  formula (\ref{red}), we make the transition to the functional integration (\ref{bv2}).
\subsection{Sturm–Liouville problem}
\label{41}
Let $\tau>0$. We consider the Sturm–Liouville problem $-\frac{1}{2}y_{ss}''(s)=\lambda y(s)$ on the interval $[0,\tau]$
with the homogeneous boundary conditions $y(0)=y(\tau)=0$.
It is known that $\lambda_n=\frac{1}{2}(\pi n/\tau)^2$ and $\psi_n(s)=\sqrt{2/\tau}\sin(\sqrt{2\lambda_n}s)$ are its eigenvalues and eigenfunctions, respectively. Green's function $g(s,t)$ for such problem is a solution of the system
\begin{equation*}
\begin{cases}
-\frac{1}{2}g_{ss}''(s,t)=\delta(s-t);\\
g(0,t)=g(\tau,t)=0,\,\,t\in[0,\tau],
\end{cases}
\end{equation*}
and has the following form (see \cite{f+8})
\begin{equation}
\label{vf}
g(s,t)=
\begin{cases}
\frac{2s}{\tau}(\tau-t),&s\leqslant t;\\
\frac{2t}{\tau}(\tau-s),&s\geqslant t.
\end{cases}
\end{equation}
In a further work its derivatives with respect to the arguments will be required. In addition to the obvious calculations, we need to determine the derivative at the break point
\begin{equation}\label{e16}
\left.\frac{d}{ds}g(s,t)\right|_{t=s}=
\left.\frac{1}{2}\frac{d}{ds}g(s,t)\right|_{t=s+0}+
\left.\frac{1}{2}\frac{d}{ds}g(s,t)\right|_{t=s-0}=1-\frac{2s}{\tau},
\end{equation}
as well to find the second derivative
\begin{equation}\label{e17}
\frac{d}{ds}\frac{d}{dt}g(s,t)=
-\frac{2}{\tau}+2\delta(s-t).
\end{equation}

\subsection{Basic notations}
\label{42}
Let $\{\eta_{\mu}\}_{\mu=1}^d$: $\eta_{\mu}\in C^{\infty}(0,\tau)$ for all $\mu\in\{1,\ldots,d\}$,  is a set of external sources. We denote  $\eta=(\eta_1,\ldots,\eta_d)$ and  assume that they satisfy the relation
\begin{equation}\label{e11}
\frac{\delta\eta_{\mu}(s)}{\delta\eta_{\nu}(t)}=\delta_{\mu}^{\nu}\delta(s-t),
\end{equation}
which is called the functional derivative.
Next, it is convenient to introduce the functional
\begin{equation}\label{e10}
b\left(\eta\right)=\frac{1}{2}\int_0^{\tau}\int_0^{\tau}dsdt\,\eta_{\mu}(s)g(s,t)\eta^{\mu}(t),
\end{equation}
whose time derivative $\tau$, taking into account the Green's function (\ref{vf}), has the following form
\begin{equation}\label{e5}
\frac{d}{d\tau}b(\eta)=\frac{1}{\tau^2}\int_0^{\tau}ds\,s\eta_{\mu}(s) \int_0^{\tau}dt\,t\eta^{\mu}(t).
\end{equation}
Also, for brevity, it is necessary to introduce an additional notation:
\begin{multline}
\label{iu}
M\left(\frac{\delta}{\delta\eta},x-y,t, \tau\right)=-f^{FS}_\mu\left(\frac{\delta}{\delta\eta(t)}+\frac{t(x-y)}{\tau}\right)\left(\frac{d}{dt}\frac{\delta}{\delta\eta_\mu(t)}+\frac{(x-y)^\mu}{\tau}\right)+\\+V\left(\frac{\delta}{\delta\eta(t)}+\frac{t(x-y)}{\tau}\right),
\end{multline}
where $t\in[0,\tau]$, $x,y\in\Omega$, and in $f^{FS}_{\mu}(x-y,y)$ and $V(x-y,y)$  the second argument is omitted, since it is not important in the proof.

\subsection{Exponential formula}
\label{43}
\begin{theorem}
\label{uy}
Let $x, y\in\Omega$, $a_k(x,y)$ are functions from (\ref{w1}), and $\tau>0$ is quite small, then, using the definitions (\ref{e11})-(\ref{iu}), we have
\begin{equation}\label{e2}
P_te^{\int_0^{\tau}dt\,M\left(\frac{\delta}{\delta\eta},x-y,t, \tau\right)}e^{b(\eta)}\bigg|_{\eta=0}= \sum_{k=0}^\infty \tau^k a_k(x,y).
\end{equation}
\end{theorem}
\noindent\textbf{Proof:} We first  consider the case without the potential. Regarding the left hand side of the formula (\ref{e2}), it is important to make an observation. Let us do the change of variables in the form $t\rightarrow \tau t$ in all integrals, then given the $\delta$ - function property $\delta(\tau t-\tau s)=\tau^{-1}\delta(t-s)$, we can go to a set of new external sources $\eta_{\mu}(\tau t)\rightarrow\hat{\eta}_{\mu}(t)$ on the interval $[0,1]$.
At the same time, to preserve the appearance of the relations (\ref{e11}), it is convenient to make substitutions on the left hand side of the formula (\ref{e2}) as
\begin{equation}
\frac{\delta}{\delta\eta_{\mu}(\tau t)}
\rightarrow
\frac{\delta}{\delta\hat{\eta}_{\mu}(t)},\,\,
\eta_{\mu}(\tau t)
\rightarrow
\frac{1}{\tau}\hat{\eta}_{\mu}(t).
\end{equation}
Thus, it is easy to verify that the degree of the parameter $\tau$ is equal to the degree of the functional $b(\eta)$. Hence the expansion is just
\begin{equation}
\label{e112}
P_te^{\int_0^{\tau}dt\,M\left(\frac{\delta}{\delta\eta},x-y,t, \tau\right)}\frac{1}{k!}b^k(\eta)\bigg|_{\eta=0}=\tau^kc_k(x,y).
\end{equation}
Let us verify that the coefficients $c_k(x,y)$ satisfy the same equations (\ref{e12}) as the coefficients $a_k(x,y)$, from which their equality will follow. The course of proof will be as follows. Let $k\in \mathbb{N}$. It is necessary to show that $\left(k+(x-y)^\mu\partial_{x^{\mu}}\right)c_k(x,y)$ can be represented as the sum of four terms
\begin{multline}\label{e14}
\partial_{x_{\mu}}\partial_{x^{\mu}}c_{k-1}(x,y)+\left(\partial_{x_{\mu}}f^{FS}_{\mu}(x-y)\right)c_{k-1}(x,y)+\\+f^{FS\mu}(x-y)f^{FS}_{\mu}(x-y)c_{k-1}(x,y)+2f^{FS}_{\mu}(x-y)\partial_{x_{\mu}} c_{k-1}(x,y).
\end{multline}
At this stage, it is convenient to note that the operator $\left(k+(x-y)^\mu\partial_\mu\right)\tau^{k-1}$ can be represented as a product of three operators $\tau^{-(x-y)^{\mu}\partial_{\mu}}\frac{\partial}{\partial\tau}\tau^{(x-y)^{\mu}\partial_{\mu}}\tau^k$. 
It is clear that the first operator $\tau^{(x-y)^{\mu}\partial_{\mu}}$ multiplies $(x-y)^\mu$ by $\tau$. Then the derivative of $\tau$ is taken and the last operator divides $(x-y)^\mu$ by $\tau$. After calculations, using the formula (\ref{e5}), the definition of the ordered exponential, and the equality
\begin{equation}
\tau^{(x-y)^\mu\partial_\mu}M\left(\frac{\delta}{\delta\eta},x-y,t, \tau\right)=M\left(\frac{\delta}{\delta\eta},x-y,t, 1\right),
\end{equation}
one can obtain that the result on the left hand side (\ref{e112}) has the form
\begin{multline}\label{e13}
P_te^{\int_0^{\tau}dt\,M\left(\frac{\delta}{\delta\eta},x-y,t,\tau\right)}\frac{1}{(k-1)!}b^{k-1}(\eta)\int_0^{\tau}ds\,\frac{s\eta_{\mu}(s)}{\tau}\int_0^{\tau}d\hat{s}\,\frac{\hat{s}\eta^{\mu}(\hat{s})}{\tau}\bigg|_{\eta=0}+\\+
M\left(\frac{\delta}{\delta\eta},x-y,\tau,\tau\right)P_te^{\int_0^{\tau}dt\,M\left(\frac{\delta}{\delta\eta},x-y,t,\tau\right)}\frac{1}{k!}b^k(\eta)\bigg|_{\eta=0}.
\end{multline}
First of all, it is necessary to pay attention to the first term. Considering that
\begin{equation}\label{e18}
f^{FS}_{\rho}\left(\frac{\delta}{\delta \eta(t)}+\frac{tx}{\tau}\right)\left(\frac{d}{dt}\frac{\delta}{\delta\eta_{\rho}(t)}+\frac{x^{\rho}}{\tau}\right)=f^{FS}_{\rho}\left(\frac{\delta}{\delta \eta(t)}+\frac{tx}{\tau}\right)\left(\frac{d}{dt}\frac{\delta}{\delta\eta_{\rho}(t)}-\frac{1}{t}\frac{\delta}{\delta\eta_{\rho}(t)}\right)
\end{equation}
due to antisymmetry, and
\begin{equation}\label{e15}
\left(\frac{d}{dt}\frac{\delta}{\delta\eta_{\mu}(t)}-\frac{1}{t}\frac{\delta}{\delta\eta_{\mu}(t)}\right)\int_0^{\tau}ds\,s\eta_{\nu}(s)=0,\,\,
\frac{\delta}{\delta\eta^{\mu}(t)}\int_0^{\tau}ds\,\frac{s}{\tau}\eta^{\nu}(s)=\partial_{x^{\mu}}\frac{tx^{\nu}}{\tau},
\end{equation}
it can be written as $\partial_{x_{\mu}}\partial_{x^{\mu}}\tau^{k-1}c_{k-1}(x,y)$.
As a result, we need to show that the second expression in (\ref{e13}) corresponds to the other three terms from (\ref{e14}). Taking into account permutation, we can assume that one end of the Green's function lies in the leftmost $M-$function. Considering the definition and the property (\ref{e17}) of the Green's function, as well as the relation (\ref{e15}), the second term (\ref{e13}) can be rewritten like
\begin{multline}
M\left(\frac{\delta}{\delta\eta},x-y,\tau,\tau\right)b(\eta)\bigg|_{\eta=0}\tau^{k-1}c_{k-1}(x,y)+\\+M\left(\frac{\delta}{\delta\eta},x-y,\tau,\tau\right)\int_0^{\tau}dt\,M\left(\frac{\delta}{\delta\eta},x-y,t,\tau\right)b(\eta)\bigg|_{\eta=0}\tau^{k-1}c_{k-1}(x,y)+\\
+2M\left(\frac{\delta}{\delta\eta},x-y,\tau,\tau\right)\int_0^{\tau}dt\,(\tau-t)\eta_{\mu}(t)\partial^\mu \tau^{k-1}c_{k-1}.
\end{multline}
The theorem follows from Lemma \ref{e6}. To prove the case with the potential, a similar reasoning can be made if we take into account the facts that after scaling the term in the exponential (\ref{e2}) with the potential is proportional to $\tau$, and that after differentiation an additional term will appear
\begin{equation}
V\left(\frac{\delta}{\delta\eta(\tau)}+(x-y)\right)
P_te^{\int_0^{\tau}dt\,M\left(\frac{\delta}{\delta\eta},x-y,t,\tau\right)}e^{b(\eta)}\bigg|_{\eta=0}.
\end{equation}
Further, from the definition of Green's function (\ref{vf}) it follows that the functional derivative $\frac{\delta}{\delta\eta(\tau)}$ in the left multiplier can be removed. In this case, the multiplication operator $V(x-y)$ will arise and the proof will be completed. $\blacksquare$

\begin{lemma}\label{e6}
Let  $\tau>0$, $x\in\mathbb{R}^d$, and $f^{FS}_{\rho}(x)$ and $b(\eta)$ taken from the definitions (\ref{e1}) and (\ref{e10}), then we have the relations
\begin{equation}
-f^{FS}_{\rho}\left(\frac{\delta}{\delta\eta(\tau)}+x\right)\left(\frac{d}{d\tau}\frac{\delta}{\delta\eta_{\rho}(\tau)}+\frac{x^\rho}{\tau}\right)b(\eta)=\partial_{x_{\mu}} f^{FS}_{\mu}(x);
\end{equation}
\begin{multline}
f^{FS}_{\rho}\left(\frac{\delta}{\delta \eta(\tau)}+x\right)\left(\frac{d}{d\tau}\frac{\delta}{\delta\eta_{\rho}(\tau)}+\frac{x^\rho}{\tau}\right)\int_0^{\tau}dt\,f^{FS}_{\sigma}\left(\frac{\delta}{\delta\eta(t)}+\frac{t}{\tau}x\right)\left(\frac{d}{dt}\frac{\delta}{\delta\eta_{\sigma}(t)}+\frac{x^\sigma}{\tau}\right)b(\eta)=\\
=f^{FS}_{\rho}(x)f^{FS\rho}(x);
\end{multline}
\begin{equation}
-f^{FS}_{\rho}\left(\frac{\delta}{\delta\eta(\tau)}+x\right)\left(\frac{d}{d\tau}\frac{\delta}{\delta\eta_{\rho}(\tau)}+\frac{x^\rho}{\tau}\right)\int_0^{\tau}dt\,(\tau-t)\eta_{\mu}(t)=f^{FS}_{\mu}(x).
\end{equation}
\end{lemma}
\noindent\textbf{Proof:} To prove the first equality, it suffices to consider an arbitrary term 
$$\left(\frac{\delta}{\delta\eta(t)}+x\right)^{\rho_1\ldots\rho_k}\nabla_{\rho_1}\ldots\nabla_{\rho_{k-1}}F_{\rho_k\rho}(y)$$
from the expansion of $f^{FS}_{\rho}(x)$ and, using the formulae (\ref{e16}) and (\ref{e18}), use the equality
\begin{equation}
\left.\left(\frac{\delta}{\delta \eta(\tau)}+x\right)^{\rho_i}\left(\frac{d}{d\tau}\frac{\delta}{\delta\eta_{\rho}(\tau)}-\frac{1}{\tau}\frac{\delta}{\delta\eta_{\rho}(\tau)}\right)b(\eta)\right|_{\eta=0}
=-\delta^{\rho_i\rho}.
\end{equation}
The second and third relations follow, taking into account (\ref{e18}), from the equalities
\begin{equation}
\left(\frac{d}{d\tau}\frac{\delta}{\delta\eta_{\rho}(\tau)}-\frac{1}{\tau}\frac{\delta}{\delta\eta_{\rho}(\tau)}\right)\left(\frac{d}{dt}\frac{\delta}{\delta\eta_{\sigma}(t)}-\frac{1}{t}\frac{\delta}{\delta\eta_{\sigma}(t)}\right)b(\eta)=2\delta^{\rho\sigma}\delta(t-\tau),
\end{equation}
\begin{equation}
\left(\frac{d}{d\tau}\frac{\delta}{\delta\eta_{\rho}(\tau)}-\frac{1}{\tau}\frac{\delta}{\delta\eta_{\rho}(\tau)}\right)\int_0^{\tau}dt\,(\tau-t)\eta_{\mu}(t)=-\delta_{\mu}^{\rho},
\end{equation}
where the property $2\int_0^{\tau}ds\,\delta(s-\tau)=1$ was used. $\blacksquare$

\subsection{Path integral}
\label{44}
Let $W_{y,x}$ be a set of continuous $\gamma:[0,\tau]\rightarrow\mathbb{R}$ paths with a beginning at the point $y$ and an end at the point  $x$. Let 
$\int_{0}^{\tau}dt\,u(t)v(t)$ be a scalar product on $C(0,\tau)$. Next, consider $\mathcal{W}_{y,x}=W_{y,x}^{\times d}$, so, if $u\in\mathcal{W}_{y,x}$, then $u=(u_1,\ldots,u_d)$ and $u_{\mu}\in W_{y,x}$ for $\forall\,\mu\in\{1,\ldots,d\}$ and the scalar product has the form $(u,v)=\int_{0}^{\tau}dt\,u_{\mu}(t)v^{\mu}(t)$. We also introduce the action
\begin{equation}
\label{red1}
S[u]=\frac{1}{4}\int_{0}^{\tau}dt\,\frac{du_{\mu}(t)}{dt}\frac{du^{\mu}(t)}{dt},\,\forall\,u\in\mathcal{W}_{y,x}.
\end{equation}
To build on the physical level of rigor it is enough to know only functional integral over $W_{0,0}$ with the action $\frac{1}{4}\int_{0}^{\tau}dt\,\left(\frac{du(t)}{dt}\right)^2$. It is known from the general theory that such quantities diverge and one of the ways to give them meaning is to use regularization by means of analytical continuation of the Riemann zeta function (see \cite{i+5,z+7,y+12}). In view of this, the integral acquires a finite value. Then, by changing the normalization of the measure, the basic formula takes the following form
\begin{equation}
\label{red}
\int_{W_{0,0}}\mathcal{D}u\,e^{-\frac{1}{4}\int_{0}^{\tau}dt\,\left(\frac{du(t)}{dt}\right)^2}=\frac{1}{\sqrt{4\pi\tau}}.
\end{equation}
\begin{lemma}
\label{ry}
Let $\tau>0$, $b$ and $S$ are defined in the formulae (\ref{e10}) and (\ref{red1}), then we have
\begin{equation}
(4\pi\tau)^{d/2}\int_{\mathcal{W}_{0,0}}\mathcal{D}u\,e^{-S[u]+(\eta,u)}=e^{b(\eta)}.
\end{equation}
\end{lemma}
\noindent\textbf{Proof:} Let $v_{\mu}(t)=\int_0^{\tau}ds\,g(t,s)\eta_{\mu}(s)$, then $u+v\in\mathcal{W}_{0,0}$ for $\forall\,u\in\mathcal{W}_{0,0}$ and one can make a shift. In this case, the statement follows from the formula (\ref{red}) and the equality $S[u+v]=S[u]+(\eta,u)+b(\eta)$. $\blacksquare$
\begin{theorem}
\label{bg}
Let $x,y\in\Omega$, $\tau>0$ is quite small, and $K$, $b$, and $S$ are defined in the formulae (\ref{po}), (\ref{e10}), and (\ref{red1}) respectively, then we have
\begin{equation}
P_te^{\int_0^{\tau}dt\,M\left(\frac{\delta}{\delta\eta},x-y,t, \tau\right)}e^{b(\eta)}\bigg|_{\eta=0}=
(4\pi\tau)^{d/2}\int_{\mathcal{W}_{0,0}}\mathcal{D}u\,e^{-S[u]}
P_te^{\int_0^{\tau}dt\,M\left(u,x-y,t, \tau\right)}
\end{equation}
and
\begin{equation}
\label{bv2}
K(x,y,\tau)=\Phi(x,y)\int_{\mathcal{W}_{y,x}}\mathcal{D}v\,e^{-S[v]}
P_te^{\int_0^{\tau}dt\,M\left(v-y,0,t, \tau\right)}.
\end{equation}
\end{theorem}
\noindent\textbf{Proof:} The first formula follows from the relation $\phi[u]=\left.\phi\left[\frac{\delta}{\delta\eta}\right]e^{(\eta,u)}\right|_{\eta=0}$ for an arbitrary polynomial $\phi$ and Lemma \ref{ry}.
The second equation follows from the formula (\ref{x1}), Theorem \ref{uy}, replacing a variable in the form $v(t)=u(t)+y+\frac{t}{\tau}(x-y)$, where $u\in\mathcal{W}_{0,0}$ and $v\in\mathcal{W}_{y,x}$, and the relation $S[u]=S[v]-\frac{1}{4\tau}(x-y)^2$. $\blacksquare$
\begin{corollary}
Let $x\in\Omega$, $\tau>0$ is quite small, then we have
\begin{equation}
\label{bv9}
K(x,x,\tau)=\int_{\mathcal{W}_{0,0}}\mathcal{D}u\,e^{-S[u]}
P_te^{\int_0^{\tau}dt\,M\left(u,0,t, \tau\right)}.
\end{equation}
\end{corollary}

\section{Conclusion}
\label{5}
As it was shown, Theorems \ref{pqs} and \ref{pq} allow us to construct diagonal elements for an arbitrary second order operator on a manifold with a flat metric and without a boundary. Suppose now that the domain under consideration $\Omega$ is endowed with a smooth Riemannian metric ($g^{\mu\nu}$). In this case, it is convenient locally, in some neighborhood $U\subset\Omega$ of a point $y$, to go to normal coordinates, which are characterized by the following conditions (see \cite{y+13}): at the point $y$ the metric tensor is $\delta_{\mu\nu}$ and for any vector $z^{\mu}$ with the condition $\delta_{\mu\nu}z^{\mu}z^{\nu}=1$ and a sufficiently small $t$ curve $x^{\mu}=y^{\mu}+tz^{\mu}$ is geodesic, and $t$ is its length.\\

\noindent Then (see \cite{f+5}) the recurrence relations for the coefficients are reduced to the form (\ref{fer}) with the only change that $\partial_{x_{\mu}}\partial_{x^{\mu}}=g^{\mu\nu}(x)\partial_{x^{\mu}}\partial_{x^{\nu}}$, where
\begin{equation}
g_{\mu\nu}(x)=\delta_{\mu\nu}-\frac{1}{3}R_{\mu\rho\nu\sigma}(y)(x-y)^{\rho\sigma}+\ldots
\end{equation}
and $R_{\mu\rho\nu\sigma}$ is the Riemann curvature tensor. Thus, it is necessary to additionally take into account the metric tensor in the second derivative (see \cite{fer1}). However, you can stay in a curved space, if you go to the exponent, ordered along the geodesic (see \cite{f+3,f+4}). In the same way, you can upgrade the diagram technique \cite{t+10} to work with coefficients $a_n(x,y)$ for arbitrary $x,y\in\Omega$. One can also use Feynman diagrams for calculations by using the path integral representation (\ref{bv2}), but loop calculations are much more complicated. \\

\noindent Practically all calculations in this work were carried out by using Fock-Schwinger gauge fixing. In particular, it played a key role in the proof of Theorem  \ref{uy}. However, the formula can be generalized to the case of an arbitrary gauge.
Let $x,y\in\Omega$ and $\omega\in\mathcal{G}$, then the gauge transformations have the form
\begin{equation}
B^{\omega}_{\mu}(x)=\omega^{-1}(x)B_{\mu}(x)\omega(x)+\omega^{-1}(x)\partial_{x^\mu}\omega(x),\,\,v^{\omega}(x)=\omega^{-1}(x)v(x)\omega(x),
\end{equation}
and for small $\tau$ the following formula is correct
\begin{equation}
K(x,y,\tau)=\omega(x)\int_{\mathcal{W}_{y,x}}\mathcal{D}u\,e^{-S[u]}
P_te^{\int_0^{\tau}dt\,(-\dot{u}^{\mu}B^{\omega}_{\mu}(u)+v^{\omega}(u))}\omega^{-1}(y).
\end{equation}
In the particular case when $\omega(x)=\Phi(x,y)$, we obtain $\omega(y)=1$, $f^{FS}_{\mu}(x-y,y)$ and $V(x-y,y)$ are from the formulae (\ref{e1}) and (\ref{w6}). It is important to note that when discussing physical phenomena it is not necessary to leave a smallness of time. For example, in physics ordered exponentials for arbitrary parameters are discussed: the Wilson line (as in (\ref{bv2})) or the Wilson loop (as in (\ref{bv9})).

\section{Acknowledgments}
The authors are grateful to A. G. Pronko for the discussion of the text.
One of the authors, A. V. Ivanov, is a Young Russian Mathematics award winner and would like to thank its sponsors and jury.
This work was supported by the Russian Science Foundation (project № 18-11-00297).

\end{document}